\documentclass[useAMS,usenatbib]{mn2e}

\usepackage{amssymb}
\usepackage{epsf}
\usepackage{bm}
\title[Self-similar pressure oscillations in neutron star envelopes
        as probes of neutron star structure ]{Self-similar pressure oscillations in neutron star envelopes
        as probes of neutron star structure }
\author[A. I. Chugunov]{A. I. Chugunov\thanks{E-mail: andr.astro@mail.ioffe.ru}
\\ Ioffe Physicotechnical Institute, St. Petersburg, Russia}

\begin{document}

\pagerange{\pageref{firstpage}--\pageref{lastpage}} \pubyear{2005}

\maketitle

\label{firstpage}

\begin{abstract}
We study eigenmodes of acoustic oscillations of high multipolarity
$l \sim 100$ -- $1000$ and high frequency ($\sim 100$ kHz),
localized in neutron star envelopes. We show that the oscillation
problem is self-similar. Once the oscillation spectrum is calculated
for a given equation of state (EOS) in the envelope and given
stellar mass $M$ and radius $R$, it can be rescaled to a star with
any $M$ and $R$ (but the same EOS in the envelope). For $l\gtrsim
300$ the modes can be subdivided into the outer and inner ones. The
outer modes are mainly localized in the outer envelope. The inner
modes are mostly localized near the neutron drip point, being
associated with the softening of the EOS after the neutron drip. We
calculate oscillation spectra for the EOSs of cold-catalyzed and
accreted matter and show that the spectra of the inner modes are
essentially different. A detection and identification of
high-frequency pressure modes would allow one to infer $M$ and $R$
and determine also the EOS in the envelope (accreted or
ground-state) providing thus a new and simple method to explore the
main parameters and internal structure of neutron stars.
\end{abstract}

\begin{keywords}
 stars: neutron -- stars: oscillations.
\end{keywords}

\section{Introduction}\label{SecIntrod}

Neutron stars can be considered as resonators where various
oscillation modes can be excited. These oscillations are attracting
much attention because, in principle, they can be used to study the
internal structure of neutron stars. Some of them (for instance,
r-modes) can be accompanied by gravitational radiation. Because
neutron stars are relativistic objects, their oscillations must be
studied in the framework of General Relativity. The relativistic
theory of oscillations was developed in a series of papers by Thorne
and coauthors
\citep{ThornPaperI,ThornPaperII,ThornPaperIII,ThornPaperIV,ThornPaperV,ThornPaperVI}.
In particular, the rapid ($\sim 1$ s) damping of p-modes with
multipolarity $l = 2$ by gravitational radiation was demonstrated by
\citet{ThornPaperIII}. An exact treatment of general-relativistic
effects is complicated, but in many cases it is possible to use the
relativistic Cowling approximation \citep{McDermott1983}. An
analysis of various oscillation modes and mechanisms for their
dissipation was carried out by \citet{McDermott1988}. Let us also
note the review paper by \citet{sterg03}, which contains an
extensive bibliography. As a rule, one considers neutron star
oscillations with low values of $l$.

Although neutron stars are objects at the final stage of stellar
evolution, they can be seismically active for many reasons. Possible
mechanisms for the generation of oscillations have been widely
discussed in the literature (see, e.g.,
\citet{McDermott1988,sterg03} and references therein). Recently,
much attention has been paid to r-modes -- vortex oscillations that
can be generated in rapidly rotating neutron stars and accompanied
by powerful gravitational radiation. In addition, oscillations can
be excited in neutron stars, for example, during X-ray bursts
(nuclear explosions in outermost layers of accreting neutron stars),
bursting activity of magnetars (anomalous X-ray pulsars and soft
gamma-ray repeaters; see, e.g., \citet{kaspi04}), and glitches
(sudden changes of spin periods) of ordinary pulsars.

In this paper we focus of high-frequency ($\sim 100$ kHz)
pressure oscillations (p-modes) with high multipolarity $l \gtrsim
100$ localized in neutron star envelopes (crusts). In our previous
paper \citep{Chug2005} we have studied these oscillations for $l
\gtrsim 500$. In that case p-modes are localized in the outer
envelope (before the neutron drip point, at densities $\rho \lesssim
4 \times 10^{11}$ g~cm$^{-3}$), where the equation of state (EOS) of
stelar matter is relatively smooth. Accordingly, the oscillation
spectrum is simple and well established.

In the present paper we extend our analysis to
p-modes with lower $l$. These oscillations penetrate
into the inner envelope of the star,
where the EOS undergoes considerable softening due to
neutronization and becomes more complicated (essentially
different for ground-state and accreted matter).
We show that the neutron drip affects strongly
the oscillation spectrum. If detected, this spectrum
would give valuable information on the EOS in neutron
star envelopes and also on global parameters of neutron
stars (their masses and radii).

\section{Formalizm}\label{SecFormal}

Following \citet{Chug2005} we study oscillations localized in a thin
neutron star envelope. It is convenient to use the approximation of
a plane-parallel layer, and write space-time metric in the envelope
as
\begin{equation}
\label{metric}
 {\rm d} s^2=c^2\,{\rm d} t^2-\,{\rm d} z^2 -R^2\,
 ({\rm d}\vartheta^2+\sin^2 \vartheta\,{\rm d}\varphi^2),
\end{equation}
where the local time $t$ and local depth $z$ are related to the
Schwarzschild time $\widetilde{t}$ and circumferential radius $r$ by
\begin{equation}
\label{Eq9}
    t= \tilde{t} \, \sqrt{1-R_{\rm G}/R},\quad
    z=(R-r)/\sqrt{1-R_{\rm G}/R},
\end{equation}
$r=R$ is the circumferential radius of the stellar surface,
$\vartheta$ and $\varphi$ are spherical angles,
$R_{\rm G}=2GM/c^2$ is the gravitation radius,
and $M$ is the
gravitational mass of the neutron star.
The metric (\ref{metric}) is locally flat and allows us to
use the Newtonian hydrodynamic equations for a thin
envelope with the gravitational
acceleration
\begin{equation}
    g=\frac{GM}{R^2\sqrt{1-R_{\rm G}/R}}.
\end{equation}
The pressure in the envelope is primarily determined by degenerate
electrons and neutrons (in the inner envelope), being almost
independent of temperature $T$. Accordingly, we can use the same
zero-temperature EOS for the equilibrium structure of the envelope
and for perturbations. Employing this EOS, we neglect the buoyancy
forces and study p-modes. The linearized hydrodynamic equations (for
a non-rotating star) can be rewritten as \citep[see, e.g., the
monograph by][]{LambBook}
\begin{equation}
\label{phi}
    \frac{\partial^2 \phi}{\partial t^2}=c_{\rm s}^2\Delta \phi
    +{\bm g}\cdot\nabla \phi,
\end{equation}
where $\phi$ is the velocity potential and $c_{\rm s}^2\equiv
{\partial P_0}/{\partial \rho_0}$ is the squared sound speed.
The velocity potential can be presented in the form
\begin{equation}
    \phi=e^{\imath\omega t}\,Y_{lm}(\vartheta,\varphi)\,F(z),
\end{equation}
where $\omega$ is an oscillation frequency, and
$Y_{lm}(\vartheta,\varphi)$ is a spherical function (see, e.g.,
\citet{Varshalovich}). An unknown function $F(z)$ obeys the equation
\begin{equation}
\label{F}
    {\frac{{\rm d} {^2F}}{{\rm d} {z^2}}}+\frac{g}{c_{\rm s}^2} {\frac{{\rm d} {F}}{{\rm d} {z}}}
    +\left(\frac{\omega^2}{c_{\rm s}^2}
    -\frac{l(l+1)}{R^2}\right)F=0.
\end{equation}

The boundary condition at the stellar surface is $F(0)=0$. It comes
from the requirement of vanishing Lagrange variation of the pressure
at the surface. The formal condition $\lim_{z\rightarrow \infty}
F(z)=0$ in the stellar interior should be imposed to localize
oscillations in the envelope. Of course, the actual variable $z$ is
finite and the real depth of oscillation localization will be
controlled in calculations. The EOS of matter in neutron star
envelopes contains a sequence of first-order phase transitions
associated with changes of nuclides with growing density. These
phase transitions are relatively weak (the density jumps do not
exceed 20 per cent). We should add boundary conditions at all phase
transitions within the envelope. These are  two well known
conditions at a plain boundary of two liquids \citep{LambBook}. The
first condition can be written as
\begin{equation}
\label{bound1}
    F^{\prime}_1(z)=F^{\prime}_2(z).
\end{equation}
It ensures equal radial velocities at both sides of
the boundary. The second condition is
\begin{equation}
\label{bound2}
    F_1=\frac{\rho_2}{\rho_1}\,F_2
    +\left(\frac{\rho_2}{\rho_1}-1\right)\,\frac{g}{\omega^2}\,F^\prime_1.
\end{equation}
It comes from the requirement of pressure continuity at the
boundary. Note, that the boundary conditions (\ref{bound1}) and
(\ref{bound2}) provide a source of buoyancy which leads to the
density discontinuity of g-modes \citep[see, e.g.,][]{McDermott1990}.

Oscillations of a plane-parallel layer
for a polytropic EOS
($P\propto \rho^{1+1/n}$, $n$ being the polytropic index)  were
studied analytically by \citet{Gough}. In this case, the squared
sound speed is $c_s^2=g\,z/n$. The solution for eigenfrequencies is
\begin{equation}
\label{omegak}
    \omega^2_k=\frac gR\,
    \sqrt{l(l+1)}\,\left(\frac{2k}n+1\right),
\end{equation}
and eigenmodes are given by
\begin{equation}
\label{Fk}
     F_k(z)=\exp\left(-\sqrt{l(l+1)}\,\frac{z}{R}\right)
            L_k^{(n-1)}\left(2\,\sqrt{l(l+1)}\,\frac{z}{R}\right),
\end{equation}
where $L_k^{(n-1)}(x)$ is a generalized Laguerre polynomial
\citep{Abramovitz}, and $k=0,1,\dots$
is the number of radial nodes.

Note, that the mode with $k=0$ does not have any radial nodes; its
properties are independent of the polytropic index $n$. This mode
corresponds to the vanishing Lagrangian variation of the density
(incompressible motion). Adding the condition $\triangle
\phi=\triangle {\bm U}=0$ to Eq.~(\ref{phi}), one can easily show that
the mode with the frequency
\begin{equation}
\label{omega0}
    \omega^2_0=\frac gR\,\sqrt{l(l+1)}
\end{equation}
and the eigenfunction $F_0(z)$, defined by Eq.~(\ref{Fk}), is the
proper mode for a wide class of EOSs. Note, that the boundary conditions
(\ref{bound1}) and (\ref{bound2}) are automatically satisfied for
this mode, and it is continuous at phase transitions. The oscillation
frequency redshifted for a distant observer is
\begin{eqnarray}
\nonumber
    \widetilde{\omega}^2_0&=&\left(1-\frac{R_{\rm g}}{R}\right) \,\frac gR\,\sqrt{l(l+1)}
    \\ \label{omega0_dist}
    &=&\frac{G\,M}{R^3}\,\sqrt{1-R_{\rm G}/R}
    \,\sqrt{l(l+1)}.
\end{eqnarray}
 The
frequency $\omega_0$ will be used to normalize eigenfrequencies of
other p-modes. The number of radial nodes $k$ will be used to
enumerate the modes.

\subsection{Self-similarity and scaling}
\label{SubSecModeScale}

Let us use the equation of hydrostatic equilibrium
${\rm d}P/{\rm d} z=\rho\, g$ and
transform Eq.\ (\ref{F}) taking the equilibrium pressure $P$ as
an independent variable,
\begin{equation}
\label{Fp}
    {\frac{{\rm d} {^2F}}{{\rm d} {P^2}}}+\frac{2}{\rho\,c_{\rm s}^2} {\frac{{\rm d} {F}}{{\rm d} {P}}}
    +\frac{1}{\rho^2}\left(\frac{\omega^2}{g^2\, c_{\rm s}^2}
    -\frac{l(l+1)}{g^2\,R^2}\right)F=0.
\end{equation}
The boundary conditions (\ref{bound1}) and (\ref{bound2}) can be
written as
\begin{eqnarray}
\label{bound1 P}
  \rho_1\,\frac{{\rm d} F_1}{{\rm d} P}&=&\rho_2\,
  \frac{{\rm d} F_2}{{\rm d}P},
  \\
\label{bound2 P}
 F_1&=&\frac{\rho_2}{\rho_1}\,F_2
    +\left(\frac{\rho_2}{\rho_1}-1\right)\,
    \frac{g^2}{\omega^2}\rho_1\,\frac{{\rm d} F_1}{{\rm d} P}.
\end{eqnarray}
Therefore, Eq.\ (\ref{Fp}) with the boundary conditions (\ref{bound1
P}), (\ref{bound2 P}) and with regularity requirement can be treated
as the equation for an eigennumber $\lambda =\omega^2/g^2$
containing the scaling parameter $\zeta=\sqrt{l\,(l+1)}/(gR)$ (with
$\zeta \approx l/(gR)$ for $l \gg 1$). Accordingly, the
eigenfrequencies can be written as
\begin{equation}
    \label{scaling}
     \omega_k^2=g^2\,g_k(\zeta)
     =\omega_0^2\,f_k(\zeta).
\end{equation}
Here, $g_k(\zeta)$ and $f_k(\zeta)$ are functions which can be
calculated numerically. They are universal for all neutron stars
with a given EOS in the envelope. The velocity potentials  $F_k$ are
also universal functions of $P$. Therefore, p-mode oscillations in
stellar envelopes are self-similar and can be easily rescaled to a
neutron star with any radius and mass. In principle, this can be
used to determine $R$ and $M$ (see Sec.\ \ref{subsec_EigenFrec}).

 \section{Numerical results}\label{SecRes}

Numerical results are presented for a ``canonical'' neutron star
model, with the mass $M_{\rm c}=1.4M_\odot$ and the radius $R_{\rm
c}=10$~km. For this model, we have $g_{\rm c}\approx 2.42\times
10^{14}$~cm~s$^{-2}$,
\begin{equation}
    \omega_0\approx 1.56 \times 10^5 \left[{l(l+1)/10^4}\right]^{1/4}
    \,\mbox{s}^{-1}
\end{equation}
and (for a distant observer)
\begin{equation}
    \widetilde{\omega}_0=\omega_0 \, \sqrt{1-R_{\rm G}/R}
    \approx 1.19 \times 10^5
    \left[ {l(l+1)/10^4}\right]^{1/4}\mbox{~s}^{-1}.
\end{equation}

Oscillation frequencies have been determined via a series of
iterative trials, checking the coincidence of the mode number
and the number of radial nodes.

\subsection{Equations of state}

We employ two models of matter in neutron star envelopes, the
accreted and ground-state matter. For the accreted matter, we use
the EOS of \citet{HZ} (HZ). It was derived by following
transformations of atomic nuclei (beta captures, emission and
absorption of neutrons, pycnonuclear reactions) in an accreted
matter element with increasing the pressure. The EOS was calculated
for the densities from $\rho=3.207\times 10^7$ g~cm$^{-3}$ to
$1.462\times 10^{13}$~g~cm$^{-3}$. For lower densities, we have
taken the matter composed of $^{56}\rm Fe$ and the EOS of degenerate
electrons with electrostatic corrections. For higher densities, we
use the EOS of the ground-state matter presented by \citet {BPS}
(BPS) because, as remarked by \citet{HZ}, the HZ EOS becomes very
similar to the BPS EOS at $\rho>10^{13}$ g~cm$^{-3}$.

We have also considered envelopes composed of the ground-state (cold
catalyzed) matter. In the outer envelope we use the EOS of
\citet{HP} (HP) and the recent EOS of   \citet{RHS}  (RHS). For the
inner envelope, we employ the EOS of \citet{NegeleVautherin}.

Phase transitions in these EOSs have been treated carefully using
the boundary conditions (\ref{bound1}) and (\ref{bound2}) at any
phase transition. For comparison, we have also employed the model of
the outer envelope
composed of ground-state matter with a smoothed composition
(the smooth composition model -- SCM). In the latter case we have
included only a large density jump at the neutron drip boundary
between the inner and outer envelopes.

\begin{figure}
    \begin{center}
        \leavevmode
        \epsfxsize=80mm \epsfbox[19 16 310 237]{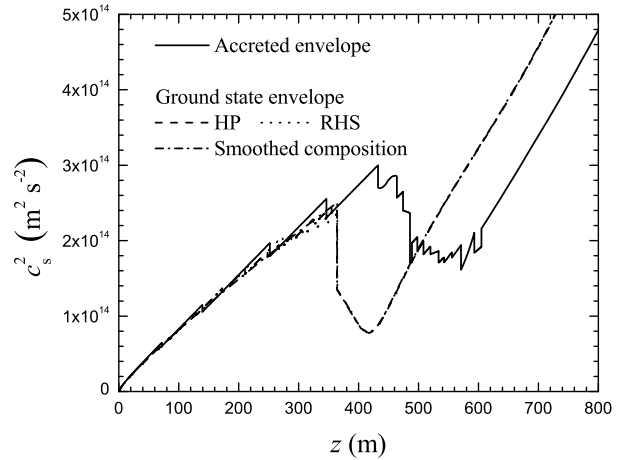}
    \end{center}
    \caption{The squared sound speed $c_s^2$ as a function of depth $z$.
    Solid line is for an accreted envelope. The dashed,
    dotted and dash-and-dotted lines
    are for the HP, RHS and SCM EOSs
    of the ground-state matter.}
    \label{fig_cs2}
\end{figure}

The squared sound speed $c_s^2$ as a function of depth $z$ for all
these EOSs is shown in Fig.\ \ref{fig_cs2}. The solid line is for
the accreted envelope; the dashed, dotted and dash-and-dot lines are
for the HP, RHS and SCM EOSs of the ground-state matter. The
different versions of the ground-state EOS show approximately the
same sound speed profiles, but the profile in the accreted envelope
is significantly different.

The depth of the accreted envelope (up to the density
$2.004\times 10^{14}$ g~cm$^{-3}$, which is the largest
density in the envelope, where the atomic nuclei are present,
for the BPS EOS) is $z\approx 1150$~m. For all models of the ground-state
matter, the largest density in the envelope has been taken $\approx
1.7\times 10^{14}$~g~cm$^{-3}$; the envelope depth is
$z\approx 985$~m.

\subsection{Eigenfrequencies}
\label{subsec_EigenFrec}

\begin{figure}
    \begin{center}
        \leavevmode
        \epsfxsize=80mm \epsfbox[20 17 289 219]{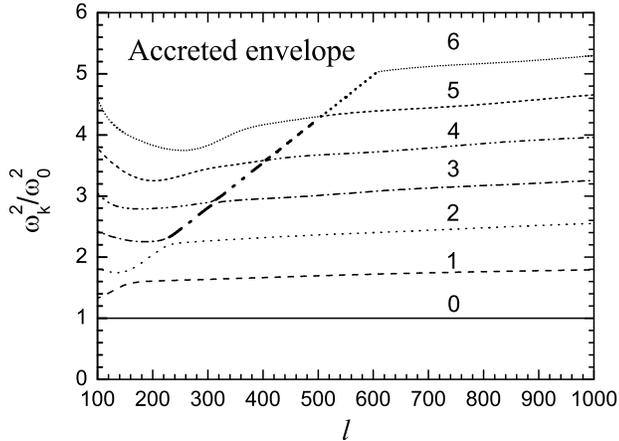}
    \end{center}
    \caption{Squared normalized
    eigenfrequencies $\omega_{k}^2/\omega_0^2$ versus
    multipolarity $l$  for the accreted envelope
    of the canonical neutron star. The
    numbers next to curves indicate the number of radial nodes.
    Thin parts of the curves correspond to the outer modes
    and (for low $l$) modes which a spread over the entire envelope,
    while thick segments refer to the inner modes.
   }
    \label{Fig_OmegAccreted}
\end{figure}

\begin{figure}
    \begin{center}
        \leavevmode
        \epsfxsize=80mm \epsfbox[20 17 289 219]{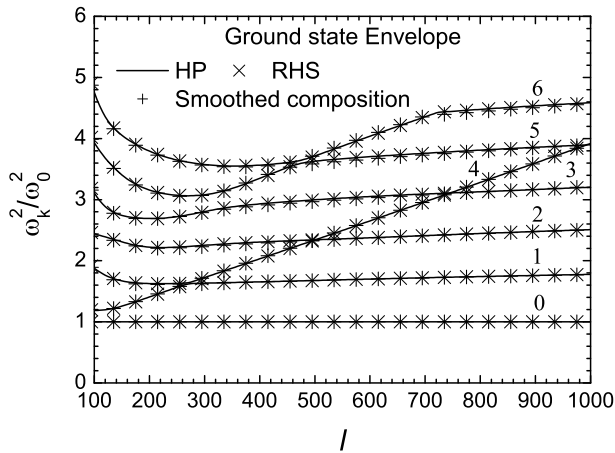}
    \end{center}
    \caption{Same as in Fig.\,\ref{Fig_OmegAccreted},
    but for the envelope composed of the ground-state matter
    (the inner modes are not emphasized).
    Lines are for the HP EOS;
    crosses `x' are for the RHS EOS;
    crosses `+' are for the SCM.}
    \label{Fig_OmegGround}
\end{figure}

Figures \ref{Fig_OmegAccreted} and \ref{Fig_OmegGround} show squares
of dimensionless eigenfrequencies $\omega_k^2/\omega_0^2$ versus
multipolarity $l$ for accreted and ground-state envelopes of the
canonical neutron star. Because of the scaling (\ref{scaling}) the
figures can be easily transformed to a star with any gravity $g$ and
radius $R$ by changing scale of the $l$ axis by a factor of
$g\,R/(g_{\rm c}\,R_{\rm c})$.

For any envelope, the modes with $l\gtrsim 300$ can be subdivided
into two groups, with a pronounced linear dependence and with a weak
dependence of $\omega_k^2/\omega_0^2$ on $l$. As will be shown in Section
\ref{subsec_EigenProfs}, the modes of the first type ({\it the inner
modes}, shown by thicker lines in Fig.\ \ref{Fig_OmegAccreted}) are
localized in the vicinity of the neutron drip point, while the modes
of the second type ({\it the outer modes}) are localized in the
outer envelope. In Figs.\ \ref{Fig_OmegAccreted} and
\ref{Fig_OmegGround} one can see a number of quasi-intersections.
When passing through a quasi-intersection point (with growing $l$),
an inner mode gains an additional radial node, but an outer mode
loses one.

Let us consider the outer modes. The eigenfrequencies are the same
(within $\sim 1\%$) for all ground-state EOSs (see Fig.\
\ref{Fig_OmegGround}). For the accreted envelope, eigenfrequencies
are larger because the EOS is stiffer. With decreasing $l$,
oscillations penetrate deeper into the outer envelope, where the EOS
is softer because electrons become relativistic, and because they
undergo beta-captures. It leads to a gradual decrease of
$\omega_k^2/\omega_0^2$. As in the model with the polytropic EOS,
given by Eq.\ (\ref{omegak}), separations between squares of
neighboring dimensionless eigenfrequencies $\omega_k^2/\omega_0^2$
are approximately constant for a fixed $l$. The weak decrease of
separations with the growth of $k$ is due to the penetration of
oscillations into deeper layers of the star, where the EOS is
softer. The latter effect is more pronounced for the ground-state
matter owing to a stronger softening of the EOS. Finally, at $l\sim
1000$ the outer p-modes are localized in the outer layers of the
outer envelope, where the matter is composed of $^{56}$Fe nuclei for
both accreted and ground-state EOSs. Accordingly, eigenfrequencies
become nearly equal.

Naturally, the oscillation frequencies of outer modes with $l\gtrsim
500$ for the ground state envelope are the same as calculated by
\citet{Chug2005}.

\begin{figure}
    \begin{center}
        \leavevmode
        \epsfxsize=80mm \epsfbox[20 17 289 219]{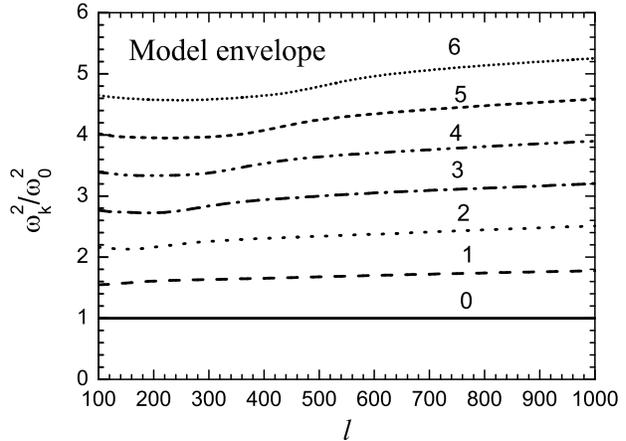}
    \end{center}
    \caption{Same as in Fig.\,\ref{Fig_OmegAccreted},
    but for a model envelope in which
    the outer envelope is composed of the ground-state matter
    and the inner envelope is composed of $^{116}$Se.}
    \label{Fig_OmegModel}
\end{figure}

To demonstrate explicitly that inner modes are caused by the neutron
drip in the inner envelope, in Fig.\ \ref{Fig_OmegModel} we present
eigenfrequencies for a model envelope without any neutron drip. Here
we employ the RHS EOS in the outer envelope but assume that the
inner envelope is composed of $^{116}$Se ions (the last element at
the outer envelope) and electron gas (no free neutrons). The
oscillation spectrum does not contain any inner modes. A small
decrease of $\omega_k^2/\omega_0^2$ for $200\lesssim l \lesssim 500$
is produced by the softening of the EOS at the bottom of the outer
envelope. The growth of frequencies at $l\sim 100$ is caused by the
penetration of oscillations into the inner envelope, where our model
EOS is polytropic (with the index $n=3$). Accordingly, oscillation
frequencies tend to the values provided by the polytropic model
(\ref{omegak}).

\subsection{Inferring $M$, $R$, and the crustal EOS
from oscillation spectrum}
\label{sect:infer}

If detected, outer modes would give
us $\widetilde{\omega}_0$, and
therefore $M\,R^{-3}\,\sqrt{1-R_{\rm G}/R}$;
see Eq.\ (\ref{omega0_dist}).
A detection of the only one fundamental mode ($k=0$) would be
sufficient to determine $M\,R^{-3}\,\sqrt{1-R_{\rm G}/R}$.
A detection of several outer modes (with different $l$ and/or $k$)
would confirm and strengthen this determination.

Our calculations show that for the inner modes
the ratio $\omega^2_{\rm in} / \omega^2_0$ is a linear
function of $l$. Using the scaling relation (\ref{scaling})
we can present this linear dependence in the form
\begin{equation}\label{InnFit}
 {\omega^2_{\rm in} / \omega^2_0}=A+B\,l, \quad B=\beta/(g_{14}R_6).
\end{equation}
Here, $g_{14}$ is the surface gravity in units
$10^{14}$~cm~s$^{-2}$, $R_6=R/10^6$ cm=$R/10$ km,
while $A$ and $\beta$ are dimensionless constants
determined by the EOS in a neutron star envelope.

For the canonical neutron star with the ground-state envelope,
we obtain $A=0.75$ and
$B=0.0032$ in the case of inner modes with lowest frequencies.
For the same star with the accreted crust we have
$A=0.65$ and $B=0.0073$. The values of $B$ allow us to determine
$\beta$. In this way we obtain
\begin{eqnarray}
 A=0.75,\quad \beta=0.0013 && {\rm for~ground~state~crust};
\label{groundfreq}
\\
 A=0.65, \quad \beta=0.0030 && {\rm for~accreted~crust}.
\label{accfreq}
\end{eqnarray}
Hence, the difference between the ground-state and accretion envelopes
is quite pronounced in oscillation spectra.

Therefore, if several (minimum two) inner modes could be detected in
addition to outer modes, their frequencies could be fitted by a
function (\ref{InnFit}) and the values of $A$ and $B$ could be
determined. An accurate determination of $A$ would enable one to
distinguish between the ground-state and accretion envelopes. The
value of $B$ would give then $g R$. Combining this $g R$ with the
value $g\,R^{-1}\,\sqrt{1-R_{\rm G}/R}$, determined from the
detection of the outer modes, one would get a simple system
of two equations for two unknowns, $M$ and $R$. Thus, a detection of
one outer mode and several inner ones could in principle enable one
to discriminate between the ground-state and accreted envelopes and
determine neutron star mass and radius.

\subsection{Eigenmodes}
\label{subsec_EigenProfs}

\begin{figure}
    \begin{center}
        \leavevmode
        \epsfxsize=80mm
    \epsfbox[20 17 289 219]{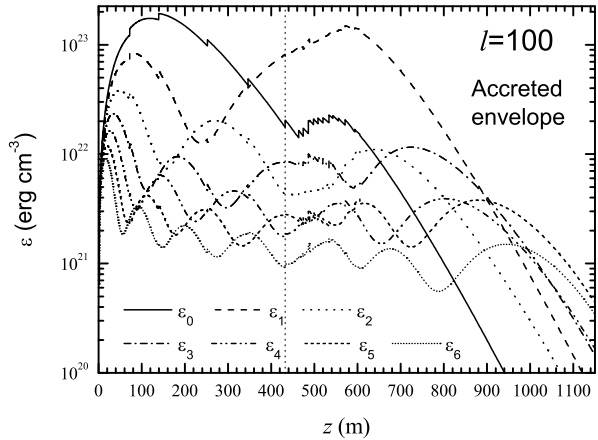}
    \end{center}
    \caption{
    Angle-averaged energy density
    of oscillations
    for modes with $l = 100$ in the accreted envelope.
    The subscript of $\varepsilon$ indicates the number of
    radial nodes. The root-mean-square amplitude of radial displacements
    at the stellar surface is 1~m.
    The vertical dotted line
    shows the boundary between the inner and outer envelopes.}
    \label{Fig_EigenAccrL100}
\end{figure}

\begin{figure}
    \begin{center}
        \leavevmode
        \epsfxsize=80mm \epsfbox[20 17 289 219]{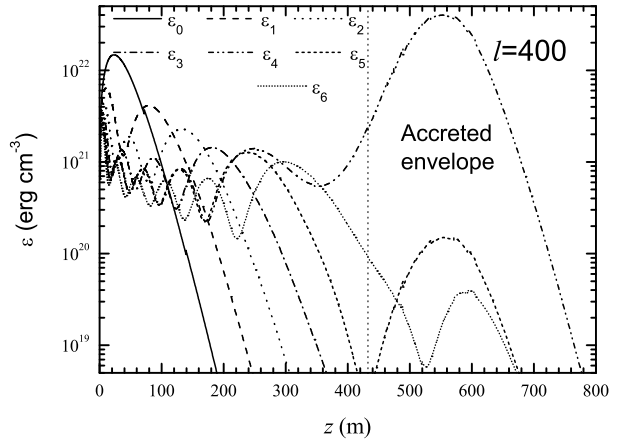}
    \end{center}
    \caption{Same as in Fig.\ \ref{Fig_EigenAccrL100}
    for modes with $l = 400$ in the accreted envelope.
     }
    \label{Fig_EigenAccrL400}
\end{figure}

\begin{figure}
    \begin{center}
        \leavevmode
        \epsfxsize80mm \epsfbox[20 17 289 219]{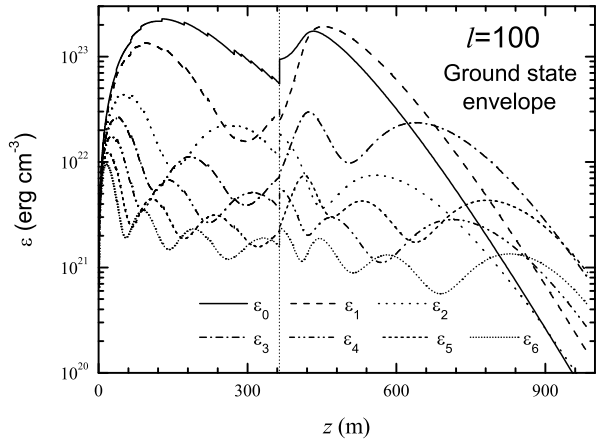}
    \end{center}
    \caption{
    Same as in Fig.\ \ref{Fig_EigenAccrL100}
    for modes with $l = 100$ in the ground-state envelope.}
    \label{Fig_EigenGrounL100}
\end{figure}

\begin{figure}
    \begin{center}
        \leavevmode
        \epsfxsize=80mm \epsfbox[20 17 289 219]{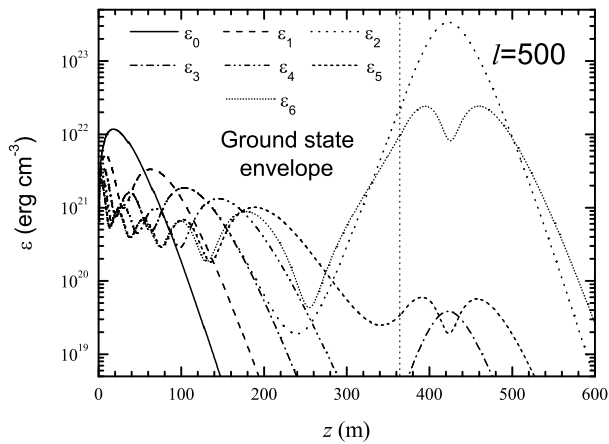}
    \end{center}
    \caption{
     Same as in Fig.\ \ref{Fig_EigenAccrL100}
     for modes with $l = 500$ in the ground-state envelope.
    }
    \label{Fig_EigenGrounL500}
\end{figure}

Figures \ref{Fig_EigenAccrL100}--\ref{Fig_EigenGrounL500}
show profiles of the angle-averaged energy density of oscillations as
a function of $z$. The root-mean-square amplitude of radial
displacements of the stellar surface has been set equal to 1~m. The
subscript of $\varepsilon$ indicates the number of radial nodes.
The vertical dotted line marks the boundary
between the inner and outer envelopes
($z\approx 432$~m for the accreted envelope,
and $z\approx 364$~m for all ground-state envelopes
of the canonical neutron star).

Figure \ref{Fig_EigenAccrL100} depicts eigenmodes with $l=100$ for the
accreted envelope. The modes are spread over the entire envelope;
their subdivision into the outer and inner modes is not obvious.
However, some traces of two mode types are visible. The mode with
one radial node (the dashed line), whose frequency belongs to the
branch of the outer modes, is primarily localized in the vicinity
of the neutron drip point. However, other modes do not demonstrate
this feature. The effects of phase transitions are relatively
small ($\sim 20$\%) and only slightly noticeable in Figure
\ref{Fig_EigenAccrL100}. They are local and do not change global
(on scales of $\gtrsim 20$~m) energy density profiles.

Figure \ref{Fig_EigenAccrL400} shows eigenmodes with $l=400$ for the
accreted envelope. This value of $l$ is very close to the
quasi-intersection point for the modes with 4 and 5 radial nodes
(see Fig.\ \ref{Fig_OmegAccreted}). The subdivision into the outer
and inner modes is clear -- the modes with $k=$0, 1, 2, 3, 5, 6
radial nodes are localized in the outer envelope, but the energy of
the mode with $k=$4 is concentrated in the outer part of the inner
envelope. This subdivision is the same as in Section
\ref{subsec_EigenFrec} (see Figure \ref{Fig_OmegAccreted}). The
energy-density profiles of the fourth and fifth modes are very
similar at $z\lesssim 250$~m, but at $z\sim 500$~m the energy
densities differ by more than two orders of magnitude! The outer
modes 'feel' the lowering of the sound speed in the outer layers of
the inner envelope (see Fig.\ \ref{fig_cs2}) and increase their
energy density in this region. However, the increase is not so large
as for the inner modes. The signatures of phase transitions are very
small ($\sim 10$\%) and hardly visible in Figure
\ref{Fig_EigenAccrL400}. They are local and do not change energy
density profiles on scales $\gtrsim 20$~m.

 Figures
\ref{Fig_EigenGrounL100} and \ref{Fig_EigenGrounL500} are plotted
for the ground-state envelope with the RHS EOS. The results for
the HP and SCM are qualitatively the same.

Figure \ref{Fig_EigenGrounL100} depicts eigenmodes with $l=100$. The
modes are localized in the entire envelope and cannot be subdivided
into the outer and inner ones. The traces of these two types of
modes are weaker than for the accreted envelope (see Fig.\
\ref{Fig_EigenAccrL100}). The signatures of phase transitions in the
outer envelope are small $\lesssim 10$\% and have scales $\sim
10$~m. They are noticeable only for the modes with a few number of
radial nodes. Many modes show large ($\sim 50$\%) jumps of the
energy density at the neutron drip point.

Figure \ref{Fig_EigenGrounL500} shows eigenmodes with
$l=500$.  This value of $l$ is close to the
quasi-intersection point for modes with $k=$2 and
3 and with $k=$5 and 6 (see Fig.\ \ref{Fig_OmegGround}).
The modes can  obviously be subdivided to the two types. The modes with
$k=$0, 1, 3, 4, 5 are localized in the outer envelope;
the energy of the modes with $k=$2 and 6 is concentrated near
the neutron drip point.  The subdivision of modes is the same as
suggested in Section \ref{subsec_EigenFrec} on the basis of
Figure \ref{Fig_OmegGround}. The energy profiles of the second and third
modes are very close for $z\lesssim 200$~m, but for  $z\sim 420$~m the
energy density differs by more than three orders of magnitude.
Qualitatively the same feature is demonstrated  by the fifth and
sixth modes. The outer modes `respond' to the lowering of the sound speed
after the neutron drip point by increasing the energy density in
this region. This increase is greater for the second and fifth
modes whose frequencies are close to the frequencies of the inner modes.
However, it
is not so large as for the inner modes. The signatures of
phase transitions in the outer envelope are extremely small ($\sim
5$\%) and are almost invisible in Figure
\ref{Fig_EigenGrounL500}. The scales of such features are $\sim
1$~m. The large phase transition at the neutron drip point
produces the signature with the same properties.

\section{Conclusions}

We have studied high-frequency pressure oscillations
which are localized in the envelopes of neutron stars composed of
the accreted or ground-state matter.

Our main conclusions are as follows.

(1) The oscillations are almost insensitive to various modifications
of the EOS for the ground-state matter (section
\ref{subsec_EigenFrec}). All EOSs we have used (HP, RHS, SCM) give
the same oscillation spectrum.

(2) The neutron drip and associated softening of the EOS in the
inner envelope do not affect strongly the spectrum of the well known
(outer) oscillation modes which are localized predominantly in the
outer envelope (section \ref{subsec_EigenFrec}).

(3) However, the neutron drip leads to the appearance of inner
oscillation modes localized mostly near the neutron drip point
(section \ref{SecRes}). The spectrum of these modes is sensitive to
the EOS in the envelope (accreted or ground-state).

(4) The p-mode oscillation problem is self-similar (in the
plane-parallel approximation). Once the problem is solved for one
stellar model, it can easily be rescaled to neutron star models with
any mass and radius (but the same EOS in the envelope; see section
\ref{SubSecModeScale}).

(5) A detection and identification of one outer mode and several
inner modes would enable one to discriminate between the
ground-state and accreted envelope and determine neutron star mass
and radius (section \ref{sect:infer}). For example, a detection of
the fundamental mode with $l=900$ at the frequency 74 kHz and of two
inner modes with $l=300$ and $l=900$ at 56 kHz and 140 kHz,
respectively, would indicate a canonical neutron star with the
ground-state envelope.

Therefore, high-frequency pressure modes are excellent tools
to explore the physics of matter in neutron star envelopes
and to determine masses and radii of neutron stars.
The oscillation frequencies could be detected by radio-astronomical
methods very precisely.
A detailed analysis of pulse shapes of some radio pulsars
reveals that oscillations with large multipolarity 
are possibly excited there \citep{Clemens2004} but their frequencies
are $\sim$30 Hz, so that they are not
high-frequency p-modes we discuss here.

A search for high-frequency p-modes could be useful.
High-multipolarity p-modes do not damp very quickly because they do
not produce any powerful gravitational or electromagnetic emission
\citep[see, e.g.,][]{Chug2005}.
 They
are robust because they are relatively independent of the thermal
state of the star, and they should not be strongly affected by
neutron star magnetic fields. The inner p-modes, localized in the
inner envelope, could be easily triggered by pulsar glitches, which
are thought to occur just in inner envelopes of pulsars.
\citet{Chug2005} studied the dissipation of p-modes localized in the
outer envelope; this dissipation is mainly produced by the shear
viscosity. It may be enhanced  by thin viscous layers near numerous
nuclear phase transitions. The viscosity in these layers can be
diffusive or turbulent. Note, that fundamental modes do not produce
viscosity layers because they pass phase transitions without
velocity discontinuities (see Sec.\ \ref{SecFormal}). Their
dissipation is not enhanced by phase transitions.

Finally, p-modes in neutron star envelopes are relatively
insensitive to the EOS and composition of neutron star cores.
However, these modes can be useful to discriminate between ordinary
neutron stars and strange stars with crust. The latter stars are
thought to contain extended cores composed of strange quark matter.
Nevertheless, a core is assumed to be surrounded by an envelope of
normal matter \citep[see, e.g.,][]{Zdunik2002}, so that a
strange star with the crust may look like an ordinary neutron star
from outside. The density of the normal matter in a strange star
does not exceed the neutron drip density. The pressure modes in the
envelopes of such stars should easily `feel' underlying dense quark
matter, and the oscillation spectrum would reflect the presence of
the quark core. We intend to consider this effect in a future
publication.

\section*{Acknowledgments}

I am grateful to D.G. Yakovlev for discussions.
This work was supported by a grant of the
''Dynasty'' Foundation and the International Center for Fundamental
Physics in Moscow, by the Russian Foundation for Basic Research
(project no.\ 05-02-16245), and by the Program of Support for Leading
Scientific Schools of Russia (NSh-1115.2003.2).

                    

\label{lastpage}

\end{document}